\documentclass[twocolumn,showpacs,showkeys,pra,footinbib]{revtex4} 
\usepackage{hyperref}  
\usepackage{graphicx}
\usepackage{dcolumn}
\usepackage{amsmath}
\begin{document}

\title{Multipole Nonlinearity of Metamaterials}
\date{\today}
\author{J. Petschulat}
\email{joerg.petschulat@uni-jena.de}
\author{A. Chipouline}
\author{A. T{\"u}nnermann}
\altaffiliation[also with: ]{Fraunhofer Institute of Applied Optics and Precision Engineering Jena, Germany.} 
\author{T. Pertsch}
\affiliation{Institute of Applied Physics,
Friedrich-Schiller-Universit{\"a}t Jena, Max Wien Platz 1, 07743,
Jena, Germany}
\author{C. Menzel}
\author{C. Rockstuhl}
\author{F. Lederer}
\affiliation{Institute of Condensed Matter Theory and Solid State
Optics, Friedrich-Schiller-Universit{\"a}t Jena, Max Wien Platz 1,
07743, Jena, Germany}
\begin{abstract}
We report on the linear and nonlinear optical response of metamaterials evoked
by first and second order multipoles. The analytical ground on which our
approach bases permits for new insights into the functionality of
metamaterials. For the sake of clarity we focus here on a key geometry, namely
the split-ring resonator, although the introduced formalism can be applied to
arbitrary structures. We derive the equations that describe linear and
nonlinear light propagation where special emphasis is put on second harmonic
generation. This contribution basically aims at stretching versatile and
existing concepts to describe light propagation in nonlinear media towards the
realm of metamaterials.
\end{abstract}
 \maketitle

\section{Introduction}
Metamaterials are attracting much scientific interest due to their unique
optical properties in terms of their tailorable effective material response and
the potential exciting applications
\cite{Pollard2009,Valentine2008,Pendry2008,Cai2007}. Most notably, the
effective magnetic response is an extraordinary property that distinguishes
metamaterials from all naturally available media at optical frequencies. It has
been shown that the effective material properties, such as the effective
magnetic and electric response, are determined by the resonant excitation of
plasmonic eigemodes, dependending on the specific metaatom geometry
\cite{Rockstuhl2006,Rockstuhl2007}. Frequently the structure and physics of
these eigenmodes is investigated by calculating the electric fields or currents
inside the metaatom by powerful
numerical techniques.\\
Here we present a more physical approach to characterize these eigenmodes by a
multipole expansion up to the second order. This model permits for an analytical treatment of the light propagation in metamaterials at the expense that it constitutes merely a simplification of the real plasmon dynamics. Nevertheless, exactly for this reason the model provides unprecedented physical insights into the functionality of metamaterials which were inaccessible before by a pure numerical treatment. In the quasi-static approach for
spherical or elliptical metal nanoparticles the fundamental plasmonic eigenmode
is sufficiently described by an electric dipole. Metamaterials, which in
general exhibit more complex geometries with an involved carrier dynamics,
require to account for higher order multipoles as e.g. the electric quadrupole
and the magnetic dipole moment representing the natural extension towards a
fully electrodynamic model \cite{Esteban2008,Han2008}. Furthermore the excitation of multipoles is frequently applied in order to explain the origin of the magnetic or the electric effective parameter response, phenomenologically. Here we present a coherent description corresponding to the way metamaterials are currently understood, namely by their multipole excitations. In particular, the simultaneous consideration of
the magnetic dipole and the electric quadrupole is required by nature, since
both occur in the same order of the multipole expansion. Though this has been
extensively discussed in the literature
\cite{Sharma2007,Cho2008,Shin2000,Guzatov2008,Petschulat2008,Pustovit2002}, the quadrupole
moment is frequently dropped.\\
Besides the linear properties that can be covered by this expansion we present
the extension of the multipole description towards the quadratic nonlinear
optical regime. Since the multipole expansion will be truncated beyond second
order terms, we focus on the study of quadratic nonlinear effects associated
with these second order multipoles. We mention that this procedure of
introducing nonlinearity is known from the early works in nonlinear optics
\cite{Bloembergen1996,Pershan1963} and is supported by several papers that
observed multipole induced nonlinear optical effects in various plasmonic
nanostructures
\cite{Kujala2007,Bethune1981,Bachelier2008,Jayabalan2008,Kim2008,Maluckov2008,Shadrivov20081,Shadrivov20082,Zharov2003}. \\
Hence, the subject of this paper is twofold. At first we derive the
required wave equations for the complete metamaterial system and compare the
 results (the dispersion relation, and the effective material
parameters) in the linear limit with numerical calculations for a
representative metaatom, i.e. the split-ring resonator (SRR). However, the
formalism we introduce is general and can be straightforwardly applied to other
geometries. The SRR was only chosen because first experiments on the second
harmonic generation were already reported in this structure; though in a
configuration amenable for nano-fabrication where the induced magnetic dipole
is non-radiating \cite{Klein2006,Klein2007}. To become specific in the second part second
harmonic generation (SHG) in a SRR is studied in detail. In order to predict the second harmonic generation by purely analytical means, we finally apply the undepleted pump approximation (UDPA) and derive the associated equations within the multipole model. We observe an excellent agreement with the numerically derived solutions which justifies the application of this approximation for further predictions.

\section{The nonlinear wave equation}
Our investigation starts with the wave equation incorporating multipoles up to
second order \cite{Petschulat2008}
\begin{eqnarray}
\Delta\mathbf{E}(\mathbf{r},t)&=&\mu_0\epsilon_0\frac{\partial^2}{\partial
t^2}\mathbf{E}(\mathbf{r},t)
+\mu_0\frac{\partial^2}{\partial t^2}\mathbf{P}(\mathbf{r},t) \label{Eq1}  \\
&-&\mu_0\nabla\cdot\frac{\partial^2}{\partial
t^2}\mathbf{\hat{Q}}(\mathbf{r},t) +\mu_0\nabla\times\frac{\partial}{\partial
t}\mathbf{M}(\mathbf{r},t). \nonumber
\end{eqnarray}

\begin{figure}
\begin{center}
\includegraphics[height=3cm]{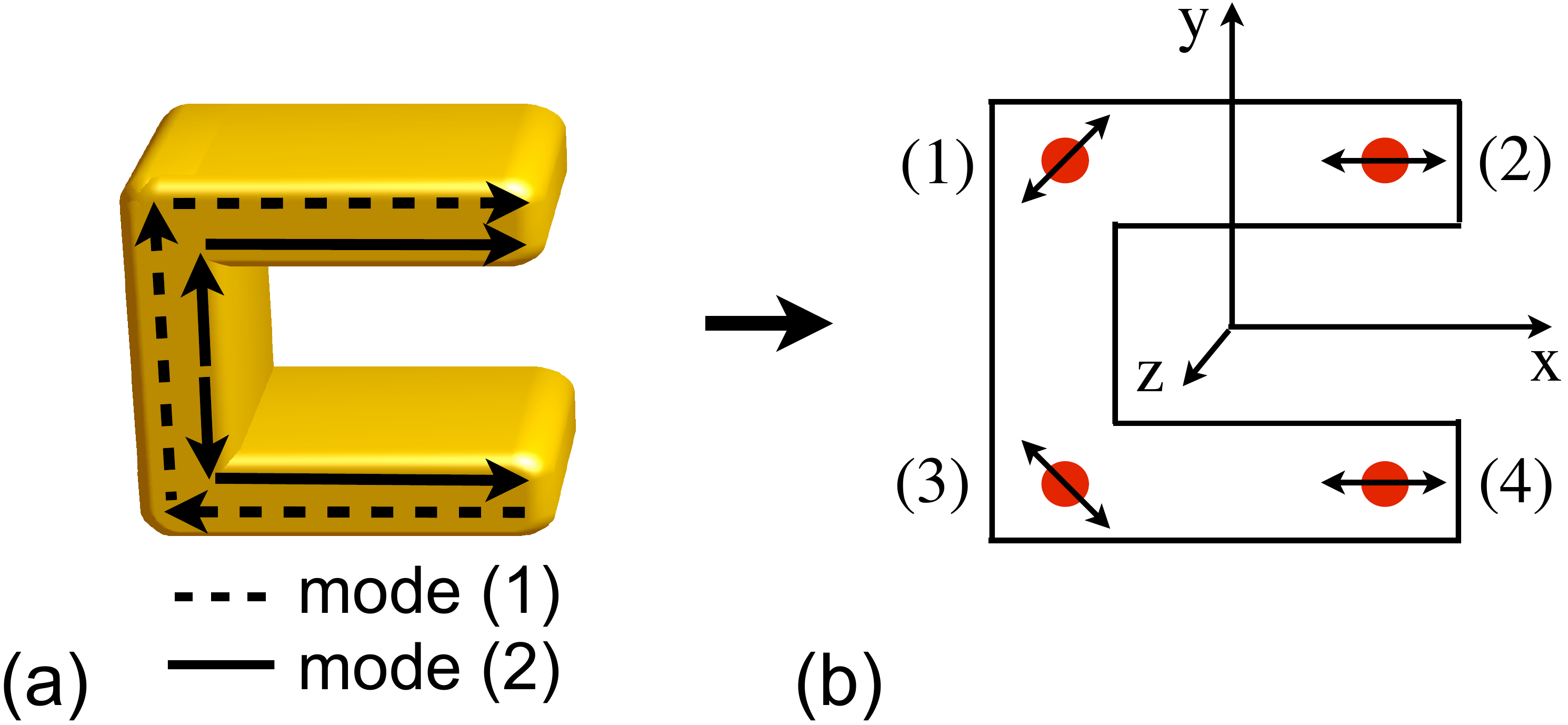}
\caption{(a) SRR metaatom and the intrinsic currents for the fundamental
electric (black solid line) and magnetic (black dashed line) mode. (b) The
associated auxiliary charge distribution (red points) with predefined degrees
of freedom (black arrows).} \label{pic1}
\end{center}
\end{figure}
Now, in terms of the plasmonic eigenmodes of interest the respective metaatom
has to be mapped onto the point multipoles: electric dipoles
$\textbf{P}(\textbf{r},t)$, magnetic dipoles $\textbf{M}(\textbf{r},t)$ and
electric quadrupoles $\mathbf{\hat{Q}}(\textbf{r},t)$. In order to observe
both, an electric and a magnetic response, the SRR is uprightly oriented
\cite{Rockstuhl2007}, see FIG.(\ref{pic1}). To cover the fundamental electric
and magnetic modes \cite{Rockstuhl2006}, sketched by the black solid and dashed
lines in FIG.(\ref{pic1}a), respectively, four auxiliary positive and negative
charges with predefined spatial degrees of freedom are required as indicated in
FIG.(\ref{pic1}b). With the knowledge about these directional constraints for
the carrier dynamics the microscopic definitions of the multipole moments can
be applied as

\begin{eqnarray}
\mathbf{P}(\mathbf{r},t)&=&\eta\sum_{l=1}^Nq_l\mathbf{r}_l(\mathbf{r},t),\label{Eq2} \\
\mathbf{M}(\mathbf{r},t)&=&\frac{\eta}{2}\sum_{l=1}^Nq_l\mathbf{r}_l(\mathbf{r},t)\times\frac{\partial}{\partial t}\mathbf{r}_l(\mathbf{r},t), \nonumber \\
Q_{ij}(\mathbf{r},t)&=&\frac{\eta}{2}\sum_{l=1}^Nq_l[\mathbf{r}_l(\mathbf{r},t)]_i[\mathbf{r}_l(\mathbf{r},t)]_j,
\nonumber
\end{eqnarray}
where the equations of motion $\mathbf{r}_l(\mathbf{r},t)$ for each charge
$q_l$ contain all information about the plasmonic eigenmodes. The parameter
$\eta$ accounts for the dipole density. Even at this early stage it can be
seen, that if $\mathbf{r}_l(\mathbf{r},t)\propto\mathbf{E}(\mathbf{r},t)$
second order multipoles will immediately evoke nonlinear contributions, since
they involve terms $\propto\mathbf{r}_l^2(\mathbf{r},t)$. For the carrier
configuration proposed here the associated terms $\mathbf{r}_l(\mathbf{r},t)$
are
\begin{eqnarray}
\mathbf{r}^+_1&=&(-x_0, y_0, 0), ~\mathbf{r}^-_1=(-x_0-\xi_1, y_0-\xi_1, 0), \label{Eq3} \\
\mathbf{r}^+_2&=&(x_0, y_0, 0),  ~\mathbf{r}^-_2=(x_0-\xi_1, y_0, 0), \nonumber \\
\mathbf{r}^+_3&=&(-x_0, -y_0, 0),  ~\mathbf{r}^-_3=(-x_0-\xi_2, -y_0+\xi_2, 0),\nonumber \\
\mathbf{r}^+_4&=&(x_0, -y_0, 0), ~\mathbf{r}^-_4=(x_0-\xi_2,
-y_0,0). \nonumber
\end{eqnarray}

In Eq.(\ref{Eq3}) the superscripts $\pm$ denote whether the position vector is
associated with a positive or a negative charge, while the argument
$(\mathbf{r},t)$ is suppressed for all $\mathbf{r}_l^\pm$. Furthermore, all
positive carriers are fixed at the positions $\pm x_0, \pm y_0$, while negative
carriers are allowed to oscillate around these sites, described by
$\xi_{1,2}=\xi_{1,2}(\mathbf{r},t)$. The distinction between the carrier
oscillations in both SRR wires is vital for realizing the two plasmonic
eigenmodes [FIG.(\ref{pic1}a)]. These carrier oscillations evoked by an
external electromagnetic field and intrinsic Coulomb interactions may be
described by a set of coupled oscillator equations as
\begin{eqnarray}
\frac{\partial^2}{\partial t^2}\xi_1+\gamma\frac{\partial}{\partial t}\xi_1+\omega_0^2\xi_1+\sigma \xi_2&=&\frac{q}{m}E_x(y+y_0,t), \label{Eq4} \\
\frac{\partial^2}{\partial t^2}\xi_2+\gamma\frac{\partial}{\partial
t}\xi_2+\omega_0^2\xi_2+\sigma \xi_1&=&\frac{q}{m}E_x(y-y_0,t).
\nonumber
\end{eqnarray}
In Eqs.(\ref{Eq4}) $\gamma$ represents the damping, $\omega_0$ the
eigenfrequency while $\sigma$ describes the coupling strength between the
carriers in the SRR arms. The physical origin of this coupling is the Coulomb
interaction of carriers in the horizontal SRR arms excited by an electric field
parallel to the arms and the carriers in the vertical arm that are excited by
the local fields of the horizontally oscillating charges, producing a current
inside the entire SRR. Moreover this coupling between two identical oscillators
results in a splitting into symmetric and anti-symmetric oscillation modes.
Substituting Eqs.(\ref{Eq3}) into Eqs.(\ref{Eq2}) the remaining multipole
moments are obtained as

\begin{eqnarray}
\mathbf{P}(\mathbf{r},t)&=&\mathbf{e}_x 2\eta q(\xi_1+\xi_2) +\mathbf{e}_y \eta q(\xi_1-\xi_2),  \label{Eq5} \\
\mathbf{M}(\mathbf{r},t)&=&-\mathbf{e}_z\frac{q\eta}{2} (x_0+2y_0)\frac{\partial}{\partial t}(\xi_1-\xi_2), \nonumber \\
Q_{xy}(\mathbf{r},t)&=&\frac{q\eta}{2}(\xi_1-\xi_2)\left[(2y_0-x_0)-(\xi_1+\xi_2)\right].
\nonumber
\end{eqnarray}
From Eqs.(\ref{Eq5}) it can be deduced that all multipoles depend either on the
sum or the difference of $\xi_1$ and $\xi_2$. Especially for the symmetric
carrier oscillation in the SRR arms ($\xi_1=\xi_2$) all second-order moments
vanish and only two identical electric dipoles parallel to the SRR arms remain
(symmetric mode). In turn, an antisymmetric oscillation ($\xi_1 =-\xi_2$)
excites both second order multipoles and a longitudinal electric dipole only
since the electric dipoles in the top and bottom SRR arms are canceling each
other (antisymmetric mode). Thus, the charge alignment chosen meets all
requirements to describe the desired plasmonic eigenmodes and their dynamics is
 determined. Decomposing the electric field into plane waves at the
 fundamental (FF) and the
 second harmonic frequency (SH) as $\mathbf{E}(\mathbf{r},t)=\mathbf{e}_xE_x(y,t)$
with
\begin{eqnarray}
E_x(\mathbf{r},t)=E_\omega e^{i(k(\omega)y-\omega t)}+E_{2\omega}
e^{i(k(2\omega)y-2\omega t)}+\text{c.c.} \label{Eq6}
\end{eqnarray}
the solutions to the oscillator equations [Eqs.(\ref{Eq4})] read as
\begin{eqnarray}
\xi_1\pm\xi_2&=&\xi^\pm_\omega E_\omega e^{i(k(\omega)y-\omega t)} \label{Eq7} \\
&+&\xi^\pm_{2\omega}E_{2\omega} e^{i(k(2\omega)y-2\omega
t)}+\text{c.c.} \nonumber
\end{eqnarray}
where the amplitudes are given by
\begin{eqnarray}
\xi^+_\omega&=&2\chi^+_\omega \cos(k(\omega)y_0) \label{Eq8}, \\
\xi^-_\omega&=&2i\chi^-_\omega \sin(k(\omega)y_0), \nonumber
\end{eqnarray}
with the introduced quasi-susceptibility
\begin{eqnarray}
\chi^\pm_\omega&=&\frac{q}{m}\frac{1}{\omega_0^2-\omega^2-i\gamma\omega\pm\sigma}.
\label{Eq9}
\end{eqnarray}
The respective equations for the SH field follow by substituting $\omega$ by
$2\omega$ in Eqs.(\ref{Eq8}). In contrast to ordinary electric dipole
interaction we observe a frequency splitting in the quasi-susceptibility
$\chi^\pm$, evoked by the two frequency degenerated eigenmodes. Now,
Eqs.(\ref{Eq5} - \ref{Eq9}) can be inserted into Eq.(\ref{Eq1}) yielding a
nonlinear eigenvalue equation with second order nonlinear source terms
\begin{widetext}
\begin{eqnarray}
\left\{\frac{\partial^2}{\partial y^2}+\frac{\omega^2}{c^2}+\omega^2\mu_0p_\omega-\mu_0\left[\omega^2q_\omega-i\omega m_\omega\right]\frac{\partial}{\partial y}\right\}E_\omega e^{ik(\omega)y}=-\omega^2\mu_0q_{\omega;2\omega,-\omega}\frac{\partial}{\partial y}\left[E_\omega^\dagger E_{2\omega}e^{i(k(2\omega)-k(\omega)^\dagger)y}\right], \label{Eq10} \\
\left\{\frac{\partial^2}{\partial
y^2}+\frac{4\omega^2}{c^2}+4\omega^2\mu_0p_{2\omega}-\mu_0\left[4\omega^2q_{2\omega}-2i\omega
m_{2\omega}\right]\frac{\partial}{\partial
y}\right\}E_{2\omega}e^{ik(2\omega)y}=-4\omega^2\mu_0q_{2\omega;\omega,\omega}\frac{\partial}{\partial
y}\left[E_\omega^2e^{i2k(\omega)y}\right]. \nonumber
\end{eqnarray}
\end{widetext}
where the following abbreviations have been used for the linear
\begin{eqnarray}
p_\omega=2\eta q\xi^+_\omega&,& m_\omega = i\omega\frac{\eta q(2y_0+x_0)}{2}\xi^-_\omega, \label{Eq11} \\
q_\omega&=&\frac{q\eta (2y_0-x_0)}{2}\xi^-_\omega, \nonumber
\end{eqnarray}
and the nonlinear multipole source terms
\begin{eqnarray}
q_{\omega;2\omega,-\omega}&=&  \frac{q\eta}{2}\left(\xi^-_\omega\xi^{+\dagger}_{2\omega}-\xi^-_{2\omega}\xi^{+\dagger}_\omega\right),\label{Eq12} \\
q_{2\omega;\omega,\omega}&=&\frac{q\eta}{2}\xi^-_{\omega}\xi^+_\omega. \nonumber
\end{eqnarray}
The exact solution to this eigenvalue equation would result in a {\it nonlinear
dispersion relation} with $k(2\omega,E_{\omega, 2\omega}), k(\omega,E_{\omega,
2\omega})$ and a fixed ratio $E_\omega/E_{2\omega}$. The left-hand side of
Eqs.(\ref{Eq10}) contains a part well-known from dipole interaction
($p_\omega$) but in addition contributions which stem from the second-order
multipole response ($q_\omega,m_\omega$). Additionally the quadrupole moment
causes a nonlinear term on the right-hand side. Interestingly, in this model
the nonlinear response of the magnetic dipole produces no nonlinear
contributions which is supported by rigorous simulations for a corresponding
SRR configuration \cite{Feth2008}. There the magnetic nonlinear contributions
have been shown to be much smaller in comparison to a convective electric
current \cite{Zeng2009} which is equivalent to the quadrupole contribution in our approach
\footnote{\it We mention that the nonlinear current as the fundamental source term
introduced in \cite{Feth2008} is equivalent to the one derived by N. Bloembergen for a
quadrupole nonlinearity \cite{Bloembergen1996}.}. Furthermore, it is mentioned
that in contrast to usual second order nonlinear optics \cite{Butcher1990} here
the nonlinear source term incorporates the first spatial derivative, induced by
the quadrupole moment.
\section{Linear Optical Properties - Effective Material Parameters}
In order to validate the predictions of the model we start with
the investigation of the linear properties. To this end the nonlinear source
terms [Eqs.(\ref{Eq12})] have been dropped which yields two decoupled linear
eigenvalue equations for the FF and the SH wave.
\begin{figure}
\includegraphics[height=7.8cm]{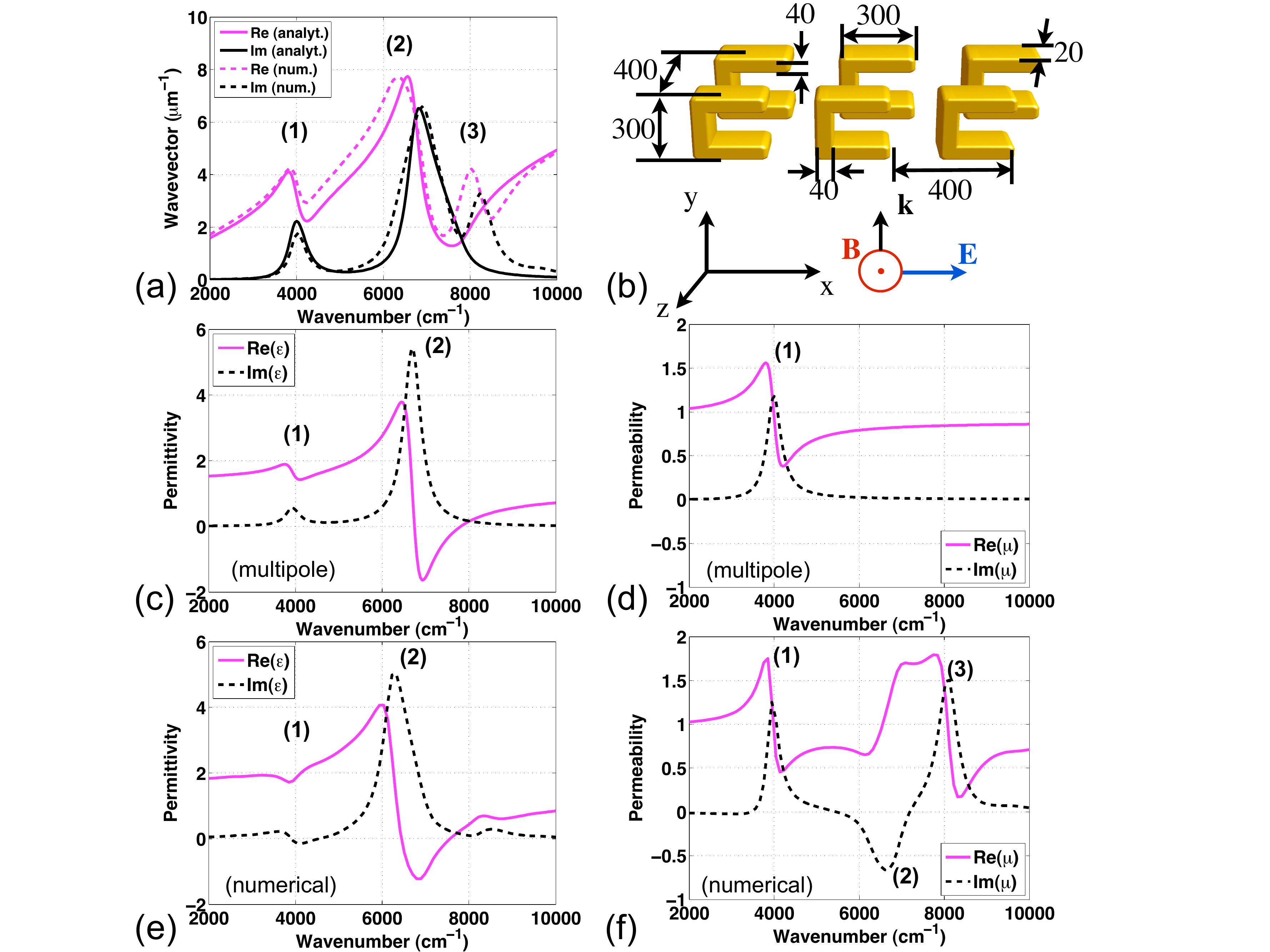}
\caption{(a) Comparison of the dispersion relation - numerical model (dashed),
analytical model (solid); (b) Single metamaterial layer (gold SRR) used for the
numerical simulation (dimensions in nm); spectral dependence of effective permittivity (c) and
effective permeability (d) from the multipole model; spectral dependence of
effective permittivity (e) and effective permeability (f) from the numerical
simulations.} \label{pic2}
\end{figure}
The obtained linear wave equations describe the field propagation in an effective medium determined by its multipolar contributions. Thus the correlated wave vector $k(\omega)$, i.e. the dispersion relation, represents a self consistent solution and contains all physical information to describe light propagation in the respective metamaterial. This can be understood in complete analogy to the exact parameter retrieval which assumes also a homogeneous wave propagation inside a metamaterial slab. 
By neglecting the the nonlinear source terms $q_{\omega;2\omega,-\omega}$ and
$q_{2\omega;\omega,\omega}$ in Eq.(\ref{Eq10}) we get the linear wave equation
\begin{eqnarray}
\left[\frac{\partial^2}{\partial y^2} +\frac{\omega^2}{c^2}+\omega^2\mu_0p_\omega-\mu_0\left[\omega^2q_\omega
-i\omega m_\omega\right]\frac{\partial}{\partial y}\right]  \nonumber \label{Eq13} \\
\times e^{ik(\omega)y}=0.
\end{eqnarray}
For plane waves their solution yields the linear dispersion relation
\begin{eqnarray}
k^2(\omega)&=&\frac{\omega^2}{c^2}\left[1+\frac{p_\omega}{\epsilon_0}-ik(\omega)\left(\frac{q_\omega}
{\epsilon_0}-i\frac{m_\omega}{\epsilon_0\omega}\right)\right].
\label{Eq14}
\end{eqnarray}
Together with the multipole definitions [Eqs.(\ref{Eq11})] this implicit equation becomes
\begin{eqnarray}
k^2(\omega)&=&\frac{\omega^2}{c^2}[1+\frac{4\eta q\chi^+_\omega \cos(k(\omega)y_0)}{\epsilon_0} \label{Eq15} \nonumber \\
&+&\frac{4y_0k(\omega)\eta q\chi^-_\omega \sin(k(\omega)y_0)}{\epsilon_0}],
\end{eqnarray} 
which has to be solved numerically. However in the subwavelength limit
$y_0<\lambda$ the trigonometric functions can be expanded
\begin{eqnarray}
\cos(k(\omega)y_0)&\approx& 1-\frac{k^2(\omega)y_0^2}{2},\label{Eq16} \nonumber \\
\sin(k(\omega)y_0)&\approx& k(\omega)y_0.
\end{eqnarray}
Thus the linear dispersion relation and hence the effective refractive index
may be explicitly written down as
\begin{eqnarray}
k^2(\omega)=\frac{\omega^2}{c^2}n^2_\text{eff}(\omega)=\frac{\omega^2}{c^2}\frac{1+\frac{4\eta q}{\epsilon_0}\chi^
+_\omega}{1+\frac{\omega^2}{c^2}\frac{2\eta q}{\epsilon_0}y_0^2\left[\chi^+_\omega-2\chi^-_\omega\right]}. \label{Eq17}
\end{eqnarray}
The remaining unknown parameters of the introduced
formalism can be obtained by fitting this dispersion relation to the
numerically calculated one for the present gold SRR geometry, shown in
FIG.(\ref{pic2}a) \cite{Rockstuhl2008}. 
As a result the two first resonances
[denoted by (1) and (2)], which are the fundamental magnetic (1) and the fundamental
electric (2) mode, are nicely covered by our dispersion relation. It can be clearly
seen that the next higher magnetic mode [resonance (3)] is not covered by the
multipole description. This would require to consider a more complex carrier
dynamic as it can be deduced from numerical simulations \cite{Rockstuhl2006}.
Moreover, the effective parameters ($\epsilon_\mathrm{eff},~\mu_\mathrm{eff}$)
follow without any further fitting \cite{Petschulat2008} and show a good
agreement in comparison with the corresponding numerically determined values
FIG.(\ref{pic2}c-f). In order to determine the effectice permittivity and permeability too we have
to consider the constitutive relations \cite{Jackson1975,Raab2005}, by which we introduce both quantities.
We start with the electric permittivity
\begin{eqnarray}
\textbf{D}(\textbf{r},t)&=&\epsilon_0\textbf{E}(\textbf{r},t)+\textbf{P}(\textbf{r},t)-\nabla\cdot\mathbf{\hat{Q}}(\textbf{r},t), \label{Eq18} \nonumber \\
D_x(y,t)&=&\epsilon_0E_x(y,t)+P_x(y,t)-\frac{\partial}{\partial y}Q_{xy}(y,t), \nonumber \\
D_\omega&=&\epsilon_0\left[1+\frac{p_\omega}{\epsilon_0}-ik(\omega)\frac{q_\omega}{\epsilon_0}\right]E_\omega ,\nonumber \\
\epsilon_\text{eff}(\omega)&\equiv&1+\frac{p_\omega}{\epsilon_0}-ik(\omega)\frac{q_\omega}{\epsilon_0}.
\end{eqnarray}
In complete analogy we derive the effective magnetic permeability and begin
with
\begin{eqnarray}
\textbf{H}(\textbf{r},t)&=&\frac{1}{\mu_0}\textbf{B}(\textbf{r},t)-\textbf{M}, \label{Eq19} \nonumber \\
H_z(y,t)&=&\frac{1}{\mu_0}B_z(y,t)-M_z(y,t), \nonumber \\
H_\omega&=&\frac{1}{\mu_0}B_\omega-m_\omega E_\omega.
\end{eqnarray}
At this point it becomes evident that the effective magnetic response is merely
evoked by the electric field \cite{Tretyakov2003}. This is a consequence of the underlying model,
because the complete carrier dynamics is driven by the electric field only.
Within the multipole expansion this carrier dynamics, i.e. the plasmonic
eigemodes are transformed into multipole moments including the magnetic dipole
moment. Thus, all moments, even the magnetic moment, are a mere consequence of
the electric field interaction with microscopic carriers. We mention that this
is occasionally erroneously interpreted, by an argumentation that the magnetic
moment is induced by the magnetic field. Our approach shows that all effects,
also magnetic ones, may be sufficiently well described by an electric field
interaction only. Nevertheless, to define the magnetic permeability it is
necessary to translate the electric field into a magnetic field which yields
\begin{eqnarray}
E_\omega&=&-\frac{\omega}{k(\omega)}B_\omega, \label{Eq20} \nonumber \\
H_\omega&=&\frac{1}{\mu_0}\left[1+m_\omega\frac{\omega}{\mu_0k(\omega)}\right]B_\omega, \nonumber \\
\mu_\text{eff}(\omega)&\equiv&\frac{1}{1+m_\omega\frac{\omega}{\mu_0k(\omega)}}. 
\end{eqnarray}
In contrast to the retrieval of the systems parameters we have fitted the disperison relation to the model, but in turn this can be performed equivalently for the effective material parameters $\epsilon_\text{eff}$ and $\mu_\text{eff}$, yielding the same physical information with respect to the required quantities.

\section{Nonlinear optical Properties - Second Harmonic Generation}
To study the nonlinear behavior induced by the fundamental modes of the present
metaatom we resort to the linear dispersion relation and treat the nonlinearity
as perturbation rather than solving Eqs.(\ref{Eq10}) exactly. As usual we rely
on the slowly varying envelope approximation (SVEA) \cite{Butcher1990}. Within
this approximation the fast spatial oscillation $\exp[ik(\omega)y]$ is
separated from a slowly varying amplitude $A(y)$, which contains all
information about the generation and depletion of the fundamental
($E_\omega=A_\omega(y)\exp[ik(\omega)y]$) and the second harmonic
($E_{2\omega}=A_{2\omega}(y)\exp[ik(2\omega)y]$).
\subsection{Exact numerical solution}
At first the solution to the wave equations incorporating the SVEA ansatz has been
performed numerically. Therefore we simplify this system by introducing the following substitutions
\begin{eqnarray}
\delta_\omega &\equiv& \frac{\omega^2}{c^2}+\omega^2\mu_0p_\omega , \label{Eq21} \nonumber \\
\beta_\omega &\equiv& \mu_0\left(\omega^2q_\omega-i\omega m_\omega\right) , \nonumber \\
\psi_{\omega;2\omega,-\omega} &\equiv& \omega^2\mu_0q_{\omega;2\omega,-\omega} , \nonumber \\
\psi_{2\omega;\omega,\omega} &\equiv& 4\omega^2\mu_0q_{2\omega;\omega,\omega} .
\end{eqnarray}

Now the eigenvalue equations take the following form
\begin{widetext}
\begin{eqnarray}
\left\{\frac{\partial^2}{\partial y^2}+\delta_\omega-\beta_\omega\frac{\partial}{\partial y}\right\}E_\omega
e^{i(k(\omega)y}&=&-\psi_{\omega;2\omega,-\omega}\frac{\partial}{\partial y}\left[E_\omega^\dagger E_{2\omega}
e^{i(k(2\omega)-k(\omega)^\dagger)y}\right] , \label{Eq22} \nonumber \\
\left\{\frac{\partial^2}{\partial y^2}+\delta_{2\omega}-\beta_{2\omega}\frac{\partial}{\partial y}\right\}
E_{2\omega} e^{i(k(2\omega)y}&=&-\psi_{2\omega;\omega,\omega}\frac{\partial}{\partial y}\left[E_\omega^2
e^{i2k(\omega)y}\right] .
\end{eqnarray}
\end{widetext}

The solution to these equations are two coupled nonlinear dispersion relations,
one for the fundamental and one for the second harmonic wave, which depend on
both fields. In order to avoid the solution of this involved system we treat
the nonlinearity as a perturbation and resort to the linear dispersion
relation. To study the effect of the nonlinear source terms we apply the slowly
varying envelope approximation where the linear fields are weighted by a slowly
varying amplitude functions $A(y)$. Replacing the constant amplitudes
$E_{\omega, 2\omega}$ by $A_{\omega, 2\omega}(y)$ we obtain upon substitution and upon neglection of the second-order derivatives (SVEA)
\begin{widetext}
\begin{eqnarray}
\frac{\partial}{\partial y}A_\omega(y)&=&-\frac{\psi_{\omega;2\omega,-\omega}}{2ik(\omega)-
\beta_\omega}\left(A_{2\omega}(y)\frac{\partial}{\partial y}A_\omega(y)^\dagger+A_\omega(y)^\dagger\frac{\partial}
{\partial y}A_{2\omega}(y)+i(k(2\omega)-k^\dagger(\omega))A_\omega(y)^\dagger A_{2\omega}(y)\right) \label{Eq23} \nonumber\\
&\times&e^{i(k(2\omega)-k(\omega)^\dagger-k(\omega))y}, \nonumber \\
\frac{\partial}{\partial y}A_{2\omega}(\omega)&=&-\frac{\psi_{2\omega;\omega,\omega}}{2ik(2\omega)-\beta_{2\omega}}
\left(2A_\omega(y)\frac{\partial}{\partial y}A_\omega(y)+2ik(\omega)A_\omega^2(\omega)\right)e^{i(2k(\omega)-k(2\omega))y}.
\end{eqnarray}
\end{widetext}

\begin{figure}
\begin{center}
\includegraphics[width=8.5cm]{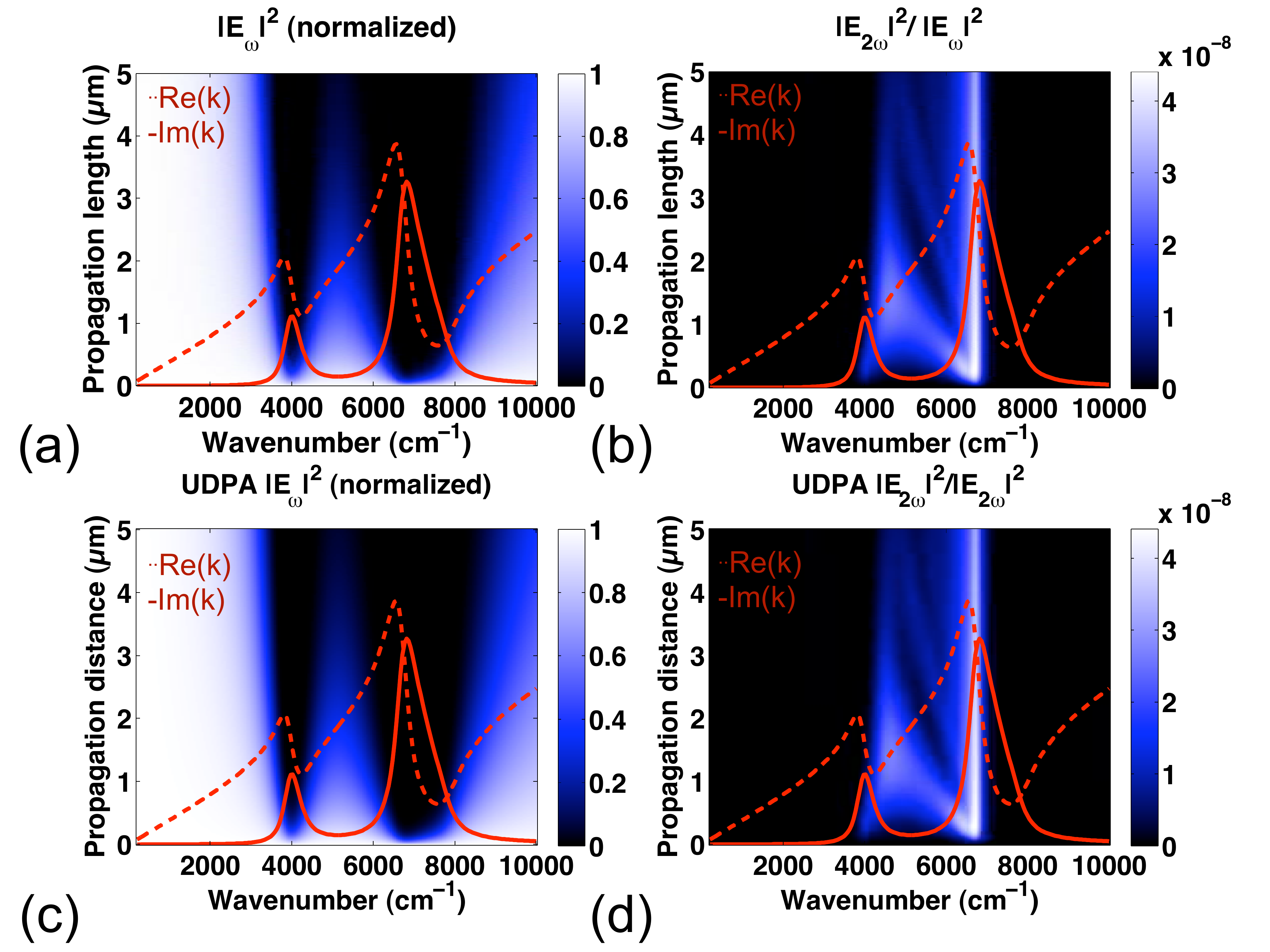}
\caption{Evolution of normalized electric field intensity for the FF (a) and
the SH (b) as a function of the wavenumber. The red lines
indicate the real (dashed) and the imaginary part (solid) of the linear dispersion relation. (c), (d) The corresponding results for the un-depleted pump approximation (UDPA).} \label{pic3}
\end{center}
\end{figure}

This final system has been solved numerically [FIG.(\ref{pic3})].
The FF wave evolution $E_\omega(y)$ is determined by the two eigenmode resonances (both
indicated by the resonances in the red lined dispersion relation), where a
strong damping is observed. For frequencies out of the spectral domain of these
resonances the FF wave propagates without excessive losses, as expected. For
the SH wave a strong contribution at the fundamental magnetic and electric
resonance can be observed. In these calculations the SHG signal originating
from the electric resonance around $\bar{\nu} = 6400~\mathrm{cm}^{-1}$ seems to
be much stronger than in the spectral vicinity of the magnetic resonance
($\bar{\nu} = 4000~\mathrm{cm}^{-1}$). This originates from the strong damping
of the SHG wave for the magnetic resonance (which propagates at $\bar{\nu}
\approx 8000~\mathrm{cm}^{-1}$), because in this spectral domain an enhanced
damping occurs due to the presence of the electric resonance. This changes
dramatically for the SH wave induced by the electric mode, since at the SH frequency
the imaginary part of $k(2\omega)$ is close to zero. Thus the second harmonic
originating from the electric resonance propagates almost without damping.
\subsection{Un-depleted pump approximation}
In order to double-check the
results and to take the weak conversion efficiency into account the undepleted
pump approximation (UDPA) has been applied \cite{Butcher1990}. Within this
approximation the fundamental wave (the pump) remains unaffected by the
generated second harmonic wave. This can be expressed by setting
$\psi_{\omega;2\omega,-\omega}$ in the first equation as well as the first
derivative of $A_\omega(y)$ in the second equation to zero. This results in
\begin{eqnarray}
\frac{\partial}{\partial y}A_{2\omega}(y)&=&\frac{-\psi_{2\omega;\omega,\omega}}{2ik(2\omega)-\beta_{2\omega}}2ik(\omega)
A_\omega^2(y)e^{i(2k(\omega)-k(2\omega)y)} , \label{Eq24} \nonumber \\
A_{2\omega}(y)&=&\frac{\psi_{2\omega;\omega,\omega}}{2ik(2\omega)-\beta_{2\omega}}\frac{2k(\omega)}{2k(\omega)-k(2\omega)}
A_\omega^2(y) \nonumber \\
&\times&\left(1-e^{i(2k(\omega)-k(2\omega)y)}\right).
\end{eqnarray}
Comparing the numerically determined electric field for the fundamental and the
second harmonic wave to those of the undepleted pump approximation [see
FIG.(\ref{pic3})] one can clearly deduce that the UDPA describes the
propagation for both waves almost exactly. Furthermore Eq.(\ref{Eq24}) permits to calculate analytically the conversion efficiency from fundamental to second harmonic intensity, e.g. for a slab consisting of a single layer of SRR's. It is important to note that for this purpose only parameters are required, that are fixed by comparison with the linear effective material response. In passing we comment that such a procedure; the determination of the nonlinear material properties based on the linear material parameters, is known as {\it Miller's delta} a well established rule in nonlinear optics \cite{Miller1964,Choy1976}.
\section{Summary} 
In summary, we have presented a self-consistent physical model that permits to
describe the linear response of metaatom geometries by their intrinsic
plasmonic eigenmodes. The occurring specific carrier dynamics have been
mimicked by an auxiliary carrier alignment interacting with the incident
radiation. The knowledge of these charge oscillations allows the application of
the multipole expansion which provides the eigenvalue equation for
electromagnetic waves propagating in such a composite metamaterial. Moreover we
have shown that the specific convective carrier oscillations together with the
quadrupole moment inherently introduce nonlinear material interactions.
Considering the SHG process, our calculations show the expected enhanced signal
both for the electric and the magnetic resonance and provide a microscopical
and physical understanding of them. For further investigations it is important
that the nonlinear response can be determined only from knowing the linear
response, which is accessible by comparing the dispersion relation or any other
effective material property to the introduced multipole model.
\\
Financial support by the Federal Ministry of Education and Research
as well as from the State of Thuringia within the Pro-Excellence
program is acknowledged.

\bibliography{ARXIV_mm_nonlinearity}
\end{document}